# UV photodetectors and field-effect transistors based on β-Ga$_2$O$_3$ nanomembranes produced by ion-beam-assisted exfoliation

*Miguel C. Pedro\*, Duarte M. Esteves, Daniela R. Pereira, Luís C. Alves, Chamseddine Bouhafs, Katharina Lorenz, Marco Peres*

M. C. Pedro, D. M. Esteves, D. R. Pereira

INESC MN, Rua Alves Redol 9, 1000-029 Lisbon, Portugal

IPFN, Instituto Superior Técnico, University of Lisbon, Av. Rovisco Pais 1, 1049-001 Lisbon, Portugal

\* miguel.cardoso.pedro@tecnico.ulisboa.pt

L. C. Alves

C2TN, Instituto Superior Técnico, University of Lisbon, Estrada Nacional 10 (km 139.7), 2695-066 Bobadela, Portugal

DECN, Instituto Superior Técnico, University of Lisbon, Estrada Nacional 10 (km 139.7), 2695-066 Bobadela, Portugal

C. Bouhafs

INESC MN, Rua Alves Redol 9, 1000-029 Lisbon, Portugal

K. Lorenz, M. Peres

INESC MN, Rua Alves Redol 9, 1000-029 Lisbon, Portugal

IPFN, Instituto Superior Técnico, University of Lisbon, Av. Rovisco Pais 1, 1049-001 Lisbon, Portugal

DECN, Instituto Superior Técnico, University of Lisbon, Estrada Nacional 10 (km 139.7), 2695-066 Bobadela, Portugal

Keywords: Gallium Oxide; Ion-beam-assisted Exfoliation; Photodetectors; Field-Effect Transistors

Abstract

β-Ga$_2$O$_3$ nanomembranes, obtained by ion-beam-assisted exfoliation, are used in the fabrication of simple metal-semiconductor-metal (MSM) structures, that are tested as photodetectors (PD) and field-effect transistors (FET). Ti/Au contacts to the membrane are found to be rectifying. However, through thermal treatment in a nitrogen atmosphere for one minute at 500°C, it is possible to modify this junction to have an ohmic behavior. An MSM PD is studied, reaching a high responsivity of 2.6×10$^4$ A/W and a detectivity of 2.4×10$^{14}$ Jones, under 245 nm wavelength illumination, and an applied voltage of 40 V. In order to better understand the behavior of the two junctions, in particular the iono/photocurrent mechanisms, an ion microprobe system is used to assess the response of these PD when excitation is localized in the different regions of the device. Finally, a depletion-mode FET is obtained, with an on/off current ratio of 7.7×10$^7$ in the linear regime, at a drain-to-source voltage of 5 V, and with a threshold voltage around −3 V. The success in obtaining FET, and most notably, MSM photodetectors, while using a simple device structure, indicates a great potential of the nanomembranes produced by ion-beam-assisted exfoliation for the development of high-performance devices.

## 1. Introduction

Gallium oxide (Ga$_2$O$_3$) is an ultrawide bandgap semiconductor, that has gained increasingly more attention over the last decade, [1] due to some of its interesting properties. Its most studied polymorph, β-Ga$_2$O$_3$, has a bandgap ($E_g$) of ~4.8 eV at room temperature, [2] and a breakdown electric field as high as 8 MV/cm, allowing potential applications in power electronics. Since solar light with wavelength below ~280 nm does not reach the Earth's surface, devices that detect only UV light with shorter wavelength are called solar-blind photodetectors (PD). [3] Wide bandgap semiconductors in general are an alternative to Si, which responds strongly to visible/infrared light (1.1 eV bandgap), thus requiring filters to be used as a UV photodiode. [3,4] Due to its wide bandgap, β-Ga$_2$O$_3$ is naturally tuned to detect these short wavelengths while being transparent to the solar spectrum. Another important advantage over other wide bandgap semiconductors such as GaN and SiC, is that large diameter bulk β-Ga$_2$O$_3$ crystals can be grown by common melt-growth techniques, allowing high-quality wafers to be manufactured at a lower cost. [5,6]

β-Ga$_2$O$_3$ has a base centered monoclinic structure and presents easy cleavage along its (100) plane. [7] This property has been explored many times for mechanical exfoliation, in the fabrication process of many different devices, such as field-effect transistors (FET) [8–14] and photodetectors, [15–20] including phototransistors. [21–29] However, conventional mechanical exfoliation, namely using the scotch-tape method, is not easily scalable and the size and thickness of the produced membranes are challenging to control. We have recently reported an innovative method allowing thin membranes of this material to be exfoliated by ion implantation. [30,31] The implantation creates defects, inducing strains and leading to the exfoliation of thin β-Ga$_2$O$_3$ layers, which self-roll into tubes. These tubes can then be easily transferred to a desired substrate and, through thermal annealing at 1000 °C, the implantation defects are recovered, the strains are relaxed, and the tubes unroll to form a nanomembrane. The result of this process is a bulk-like quality β-Ga$_2$O$_3$ nanomembrane placed on a substrate of choice.

In this work, simple devices based on β-Ga$_2$O$_3$ membranes, obtained through ion-beam-assisted exfoliation of undoped commercial single crystals were fabricated and tested to assess their potential for the development of different devices. In order to achieve that, the nature of the membranes' contact to metal was studied and metal-semiconductor-metal (MSM) structures were tested as photodetectors and field-effect transistors. Figure 1 shows a schematic and an image of one of these structures.

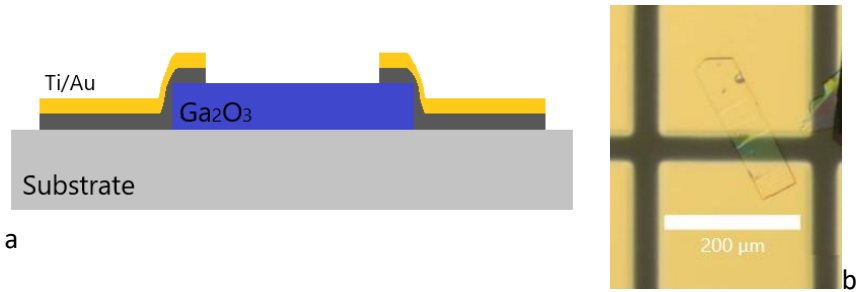

*Figure 1 – Metal-semiconductor-metal (MSM) structure: **a** – Schematic cross-section; **b** – Optical microscopy image showing the Ga$_2$O$_3$ flake, the Ti/Au areas sputtered through a ~100 mesh grid mask and the grid shadow regions (see section 4).*

## 2. Results
### 2.1. Contact optimization

When fabricating these devices, one aspect that had to be taken into account and optimized has to do with the properties of the contacts. From a device designing perspective, it is interesting to be able to achieve both ohmic and Schottky contacts. Ti has been reported as one of the best metals to make ohmic contacts to Ga$_2$O$_3$. [32] It has been observed, however, that these contacts exhibit Schottky behavior as deposited and only upon annealing do they become ohmic. Most works found an annealing at temperatures between 400 and 500 °C for 1 min in N$_2$ atmosphere to be optimal for ohmic behavior. [32–35] This is attributed to the formation of an thin intermediate interfacial Ti-TiO$_x$ layer. [34] Annealing at higher temperatures leads to the degradation of ohmic properties, [32,35] attributed to an expansion of the same interfacial layer. [35]

Since the type of contact is important for the device's behavior, the annealing temperature was optimized for our particular devices. The characteristic *I-V* curves for one of these devices are shown in Figure 2a as deposited and after annealing at temperatures ranging from 200 to 800 °C. It can be seen that these curves are not linear for the as deposited contacts, nor after annealing at 200 and 300 °C. The curve corresponding to 400 °C is already progressing towards a linear behavior, but the most ohmic behavior is found for annealing at 500 °C, similarly to what is reported in literature. After 600 °C and, more noticeably, after 700 °C annealing, a return of the contacts' properties to a Schottky-like behavior is visible, causing a non-linear *I-V* curve and a decrease in the current.

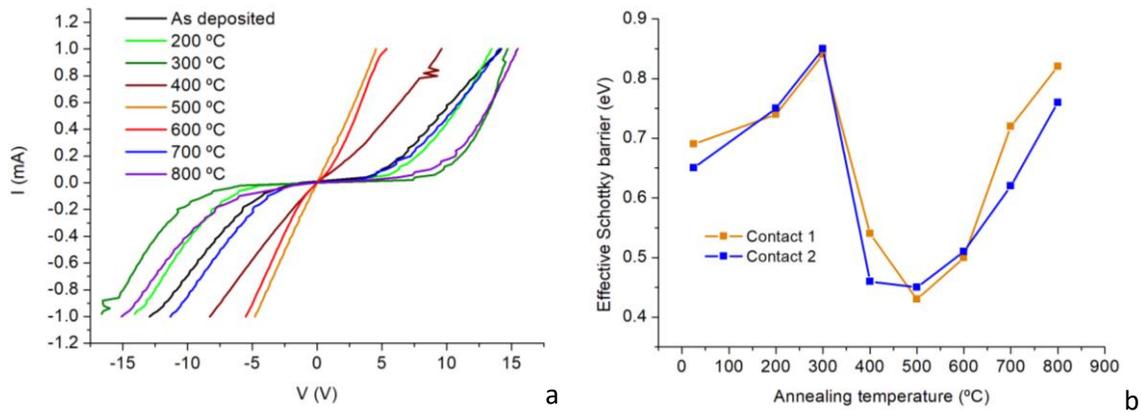

*Figure 2 – **a** – I-V curves of the MSM structure in the as deposited condition and after successive annealing steps for temperatures ranging from 200 °C to 800 °C; **b** – Evolution of the effective Schottky barrier values obtained from PKUMSM fits after each annealing step. The subscripts 1 and 2 are used to distinguish the two contacts of the device. The values for the unannealed devices are plotted at 25 °C in the temperature axis. The fits themselves are presented in the supporting informations.*

The shape of these curves corresponds very well to a general model that considers a back-to-back Schottky configuration and was proposed by Z. Y. Zhang et al.. [36] In this model, the applied voltage across an MSM device is distributed between three components: the two back-to-back Schottky barriers and the semiconductor structure in between. Thermionic emission is considered for the forward-biased contact, while thermionic-field emission is considered for the one under reverse bias. [37] According to this model, two regions may be observed in an *I-V* curve: a region at low bias, where the voltage drop at the reversely-biased contact dominates; and a region at high bias, in which the voltage drop across the semiconductor structure dominates. As a result, the low-bias region is characterized by a low current, and a width related to the Schottky barriers' height; while in the high-bias region the current increases linearly, with a slope that corresponds to the conductance of the semiconductor membrane. These distinct regions are observed very clearly in some of the curves in Figure 2a. The Schottky barrier heights in both metal/$Ga_2O_3$ interfaces are presented in Figure 2b and were independently extracted by fitting these *I-V* curves using the program PKUMSM (Peking University MSM) [38] and the corresponding fits are shown in Figure S1 in the supporting material. These values are similar between the two contacts, indicating a quite symmetrical device, which is noticeable in the *I-V* curves. As can be seen in Figure 2b, the lowest barriers are found at 500 °C, as it would be expected, since the *I-V* curve is the closest to linear after annealing at this temperature.

### 2.2. Photodetector

In order to test the nanomembrane MSM structures as photodetectors, these were fabricated on c-sapphire substrates and they were not subjected to annealing after deposition. As such, they maintain the Schottky barriers and respective depletion regions, which are essential to some photoresponse mechanisms. [39–43] Various devices were illuminated by LEDs of two different wavelengths: 245 nm (LED245); and 365 nm (LED365). The *I-V* curves for a new, representative device, henceforth referred to as device PD-A, in dark conditions and illuminated by each of these light sources are presented in Figure 3a and Figure 3b. One thing to note is that, unlike the case shown in Figure 2a, the dark current seems to begin to saturate at high bias. For lower bias, the behavior in the dark follows the previous model and the Schottky barriers were calculated to be 0.75 eV and 0.39 eV, different values for the two metallic contacts, reflecting a more asymmetrical behavior.

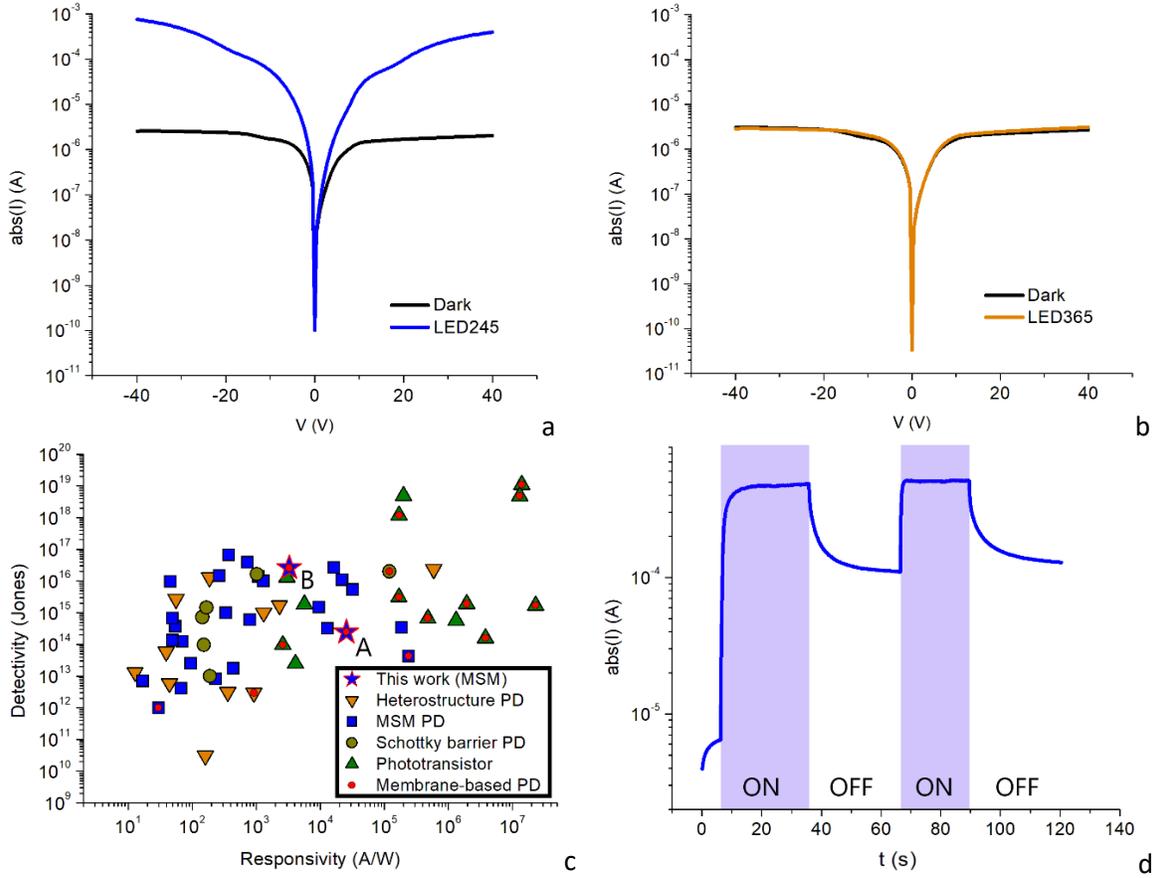

Figure 3 – I-V curves in logarithmic scale for device PD-A in the dark and under illumination from: **a** – the 245 nm wavelength LED; **b** – the 365 nm wavelength LED. **c** – Comparison of $Ga_2O_3$ photodetector devices reported in the literature, as a function of responsivity and detectivity. It divides the devices between MSM photodetectors, [18,19,40,44–65] Schottky barrier photodetectors, [17,66–70] phototransistors, [21–29,71–76] and heterostructure photodetectors. [20,77–86] Devices employing $Ga_2O_3$ membranes/flakes are marked with a red dot [17–29]. The devices discussed in the present paper (PD-A and PD-B) are highlighted with stars. Keep in mind that different conditions, such as bias voltage and light source irradiance, are used in different works. **d** – Representation of the transient measurements of the current across device PD-A, as LED245 is turned on and off. A bias V=−40 V is applied across the device.

Comparing the *I-V* curves, it is clear that the device shows a response when illuminated with above bandgap light of 5.08 eV (245 nm LED), while no clear response is observed when the device is illuminated with below bandgap light of 3.4 eV (365 nm LED), as would be expected for a material with high crystalline quality.

For device PD-A, the highest values for responsivity (*R*), photo-to-dark current ratio (*PDCR*) and detectivity (*D\**) (see the definition of these figures of merit in section 4.2) were measured at $V = -40$ V, with $R = 2.6 \times 10^4$ A/W, $PDCR = 2.9 \times 10^2$ and $D^* = 2.4 \times 10^{14}$ Jones. The responsivity, in particular, is quite high, as a result of the photocurrent almost reaching $I_{photo} = 0.75$ mA, for a detector area of $7.4 \times 10^3$ μm². Such a high responsivity indicates quite a high photoconductive gain and, in fact, for the present conditions the external quantum efficiency (*EQE*) is $1.3 \times 10^5$. When looking at the detectivity and *PDCR*, these are not as high for this device as they could potentially be, resulting from the high dark current of this device ($I_{dark}$ >1 μA). Other devices fabricated along with this one, under the same conditions, displayed a lower dark current, resulting in larger *D\** and *PDCR*. A notable example is a device, henceforth referred to as device PD-B, with higher detectivity, displaying, at *V* = −40 V, a lower *R* = 3.3x10$^3$ A/W, but *PDCR* = 2.6x10$^7$ and *D\** = 2.6x10$^{16}$ Jones. In section 2 of the supporting information, there is a short analysis of this device, along with a comparison between its *I-V* curves and those of device PD-A (Figure S2). There, the values of *R*, *PDCR* and *D\** as a function of the bias are also shown for both devices on Figure S3. This variability between devices stresses the need to control $I_{dark}$. This requires a better control of the contact fabrication since inhomogeneities, e.g. due to an inhomogeneous distribution of interface defects, can lead to increased leakage currents. [87]

These two devices compare quite well with others reported in the literature, in terms of *R* and *D\**, as can be seen in Figure 3c. The two devices reported here are near the best among the MSM $Ga_2O_3$ photodetectors, but are still behind many membrane-based PDs, in particular, those employing phototransistor structures, which allow $I_{dark}$ to be minimized using a gate voltage. In reality, reports of devices with high responsivities and gain are very common for all these structures, and that is true for devices based on bulk $Ga_2O_3$, on $Ga_2O_3$ films, and on $Ga_2O_3$ exfoliated membranes, for varying crystalline quality; a very common downside, however, are the long rise and decay times measured on those devices. [88]

Regarding response speed, and considering the transient measurements, such as the one presented in Figure 3d for device PD-A, it can be seen that after the illumination is turned on, the device reacts and the current increases to $I_{photo}$. However, when the LED is turned off, the current does not decrease to dark current values in the same time scale. These results clearly suggest that our devices suffer from a quite significant persistent photoconductivity (PPC).

A. Y. Polyakov et al. [88] report that high photoconductive gain and long response times in MSM photodetectors are mainly attributed to two factors. The first, shared with the Schottky barrier PD structures, has to do with the current crossing the reversely-biased Schottky diode. Upon illumination, the photogenerated holes can be trapped in the depletion region. This increases the positive charge concentration, thus decreasing the effective Schottky barrier, which in turn allows more current across the device. [41,89,90] The hole traps responsible for this trapping may be deep acceptors, likely related to Ga vacancies. [88,89,91] The second factor that leads to high gain is related to hole trapping in the conductive channel or at its surface. Although electrons and holes are formed in pairs, hole trapping (or spatial separation of holes and electrons due to surface band bending and the formation of a surface depletion region) will lead to a high $\Delta e/\Delta p$ ratio, i.e. the ratio between excess electrons and excess holes. If the concentration of trapped holes is high, a high concentration of electrons is present in the conduction channel, which will contribute to the photoconductivity and result in high gain. [92–96] This mechanism has also been proposed for other oxide nanostructures such as ZnO and $MoO_3$, where adsorbed oxygen is bonded with surface oxygen vacancies sites, resulting in surface band bending. [97,98] Upwards band bending at air-cleaved β-$Ga_2O_3$ (100) surfaces has also been reported previously. [99] These two gain mechanisms, both at the contacts and at the channel, also explain PPC since the electrons are not present in the depletion regions for long enough times for recombination to take place, and the main mechanism for holes to escape their traps would in theory be by thermal excitation. It is sometimes possible to accelerate this process by applying a forward bias [100] or a different gate voltage, [27,101] so that the electrons are able to reach those regions and recombination can happen.

In order to get more insight in the response mechanisms of this device (PD-A) with an MSM structure, a nuclear microprobe system was used to focus a 2.0 MeV H$^+$ beam on the two interfaces metal-semiconductor identified as 'A' and 'B' in Figure 4a and on the region between the two contacts (middle region) while measuring the *I-V* characteristics of the device in these conditions. It is worth highlighting here the particular interest of device PD-A in carrying out this study with spatial resolution, considering the differences in the work function of its two contacts, with $q\varphi_A$ = 0.75 eV and $q\varphi_B$ = 0.39 eV.

In order to understand the interaction of the beam with the sample, Figure 4b presents the values of the energy loss of the H$^+$ ions as a function of the depth in the sample calculated using SRIM. [102] Comparing the electronic and nuclear components of the energy loss, it is clear that the interaction within the device will be mainly electronic (i.e., interactions between the beam and the electrons of the target atoms), creating electron-hole pairs, which is why the effect of the beam can be compared to that of the UV LED. The nuclear interactions (i.e., interactions between the beam and the target's nuclei) should only amount to important energy losses deeper in the substrate, meaning that their effect on the device's behavior by producing defects (such as vacancies or interstitials) should be reduced. This effect seems to still exist, however, since after prolonged exposition to the ion beam, devices eventually become resistive, possibly due to an accumulation of defects that act as deep levels and pin the Fermi level. [103]

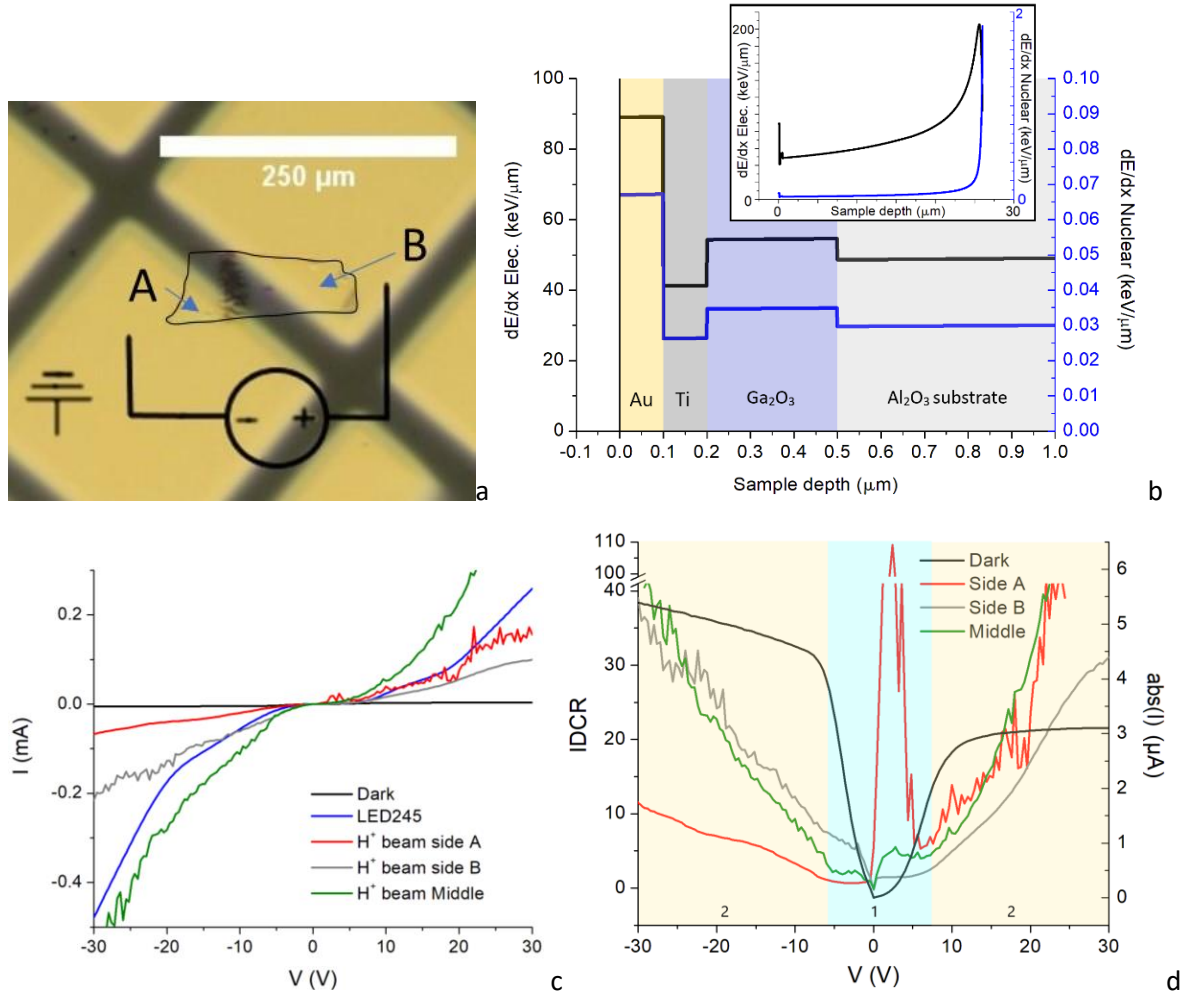

Figure 4 – *a* – Optical microscopy image of device PD-A, with an outline around the area occupied by the membrane, with the two contacts, A and B, labelled and the polarity during the measurement also represented; *b* – Curves, obtained by a SRIM [102] simulation, showing the energy loss (dE/dx) of the protons, reaching the surface at 2.0 MeV, as a function of the depth they reach in the sample, by electronic (black, left scale) and by nuclear (blue, right scale) interactions. These curves are presented for the more superficial 1 µm, with regions shown for the Au (100 nm) and Ti (100 nm) deposited layers, the $Ga_2O_3$ membrane (300 nm), and the sapphire substrate. In the inset, these curves are shown to their full extent in depth, that is, until the $H^+$ ions have deposited their energy in its entirety; *c* – I-V curves for device PD-A in the dark, under illumination by the 245 nm wavelength LED, and under a 2.0 MeV $H^+$ beam directed at side A, at side B, and at the membrane between the contacts; *d* – I-V curve for the dark current (right scale) and IDCR for the $H^+$ beam directed at the different parts of device PD-A (left scale). Two bias regions of different behavior, 1 and 2, are highlighted in blue and yellow.

The *I-V* curves corresponding to the device in the dark, under illumination and for the $H^+$ beam directed at its three regions are presented in Figure 4c. Note that the fluctuations visible in all curves are due to the fluctuation of the ion beam current, which means that the device has a response speed high enough to be sensitive to these. These curves show that the ionoresponse when the $H^+$ beam is directed at the center of the membrane is quite similar to the photoresponse to the UV LED. Focusing on the cases in which the beam was pointed at the contact regions, it is important to remember that positive bias means that Schottky diode A is in reverse bias, while for negative bias Schottky diode B is in reverse bias. For any applied voltage, the response is higher for excitation of the contact that is reversely-biased. The iono-to-dark current ratio (*IDCR*) for the beam focused on the two sides and on the middle of the device is plotted in Figure 4d, as a function of the applied bias. Interestingly, not only can a very distinct response be observed depending on where the beam is pointed, but two regions with different behavior, 1 and 2, as identified in the figure, can be found. A high response is observed in region 1 if the beam is pointed at the reversely-biased contact, while the response is very weak otherwise. This makes sense if we attribute the response to the creation of trapped holes in the depletion region of the reversely-biased contact, which is more efficient if the excitation is happening at that region. This effect is especially noticeable on side A, where *IDCR* has very tall peak, since the effective barrier for that contact is much higher than for contact B, resulting in a low dark current region. Since the effective barrier is decreased under the beam, there is not a low ionocurrent region and *IDCR* is very high. Since for side B the dark current is already growing linearly, *IDCR* does not form a peak, even though it is still higher when the contact is reversely-biased. For side A too, it would be expected that for higher bias the *IDCR* would gradually decrease, since the dark current itself also goes up. This is indeed observed but only within region 1. This is because in region 2 the dark current is reaching saturation, but not the photo- or ionoresponse, which means that the *IDCR* starts increasing again. Interestingly, in this region, the device also responds to the $H^+$ beam if it is pointed at the middle region or even at the directly-biased contact. Hole diffusion lengths in $Ga_2O_3$ have been reported to be in the order of 350-400 nm, so efficient hole diffusion to the depletion region of the Schottky contact is unlikely [104]. This may suggest that the other mechanism is dominant here, namely, hole trapping at defects in the conductive channel or at the surfaces.

Through this microprobe study, the presence of response mechanisms associated with both the reversely-polarized contact and the channel region can be observed for device PD-A. These results also suggest that the effects at the contact may not the dominant for the high biases at which the highest *R*, *D\** and *PDCR* were obtained using the LED, and instead the effects of light on the channel across the membrane are likely the most important contribution to the photoresponse in this device.

2.3.       Field-effect transistor

The MSM devices fabricated on $Si/SiO_2$ substrates were tested as gallium-oxide-on-insulator (GOOI) FET, as shown in Figure 5. At first, unannealed devices with this structure were electrically characterized. Representative output curves, showing the measured current ($I_{ds}$) as a function of the drain-to-source voltage ($V_{ds}$), are presented in Figure 5a for a device with width $w$ = 105 µm. The figure depicts these curves only for positive $V_{ds}$, which closely follow a typical FET behavior. Since the contacts are Schottky-type, there is a very small low current region at low bias, followed by the linear and saturation regions for increasing positive bias, that are typical of an n-type FET. From the output curves, the maximum drain current per unit width ($I_{ds,max}/w$) was measured, for $V_g$=40 V, to be only 0.91 mA/mm. The same device was then annealed at 500 °C, in order to create ohmic contacts. The same measurements were repeated, yielding the output curves presented in Figure 5b. It is observed that the low current region at low voltage is now absent due to the reduction of the effective Schottky barriers, as previously observed in section 2.1. The current is now nearly ten times higher, with $I_{ds,max}/w$ = 8.0 mA/mm, for $V_g$=40 V. This value can be further increased by adequately doping $Ga_2O_3$. [10] Another consequence of annealing is that saturation is only reached at a larger bias, however.

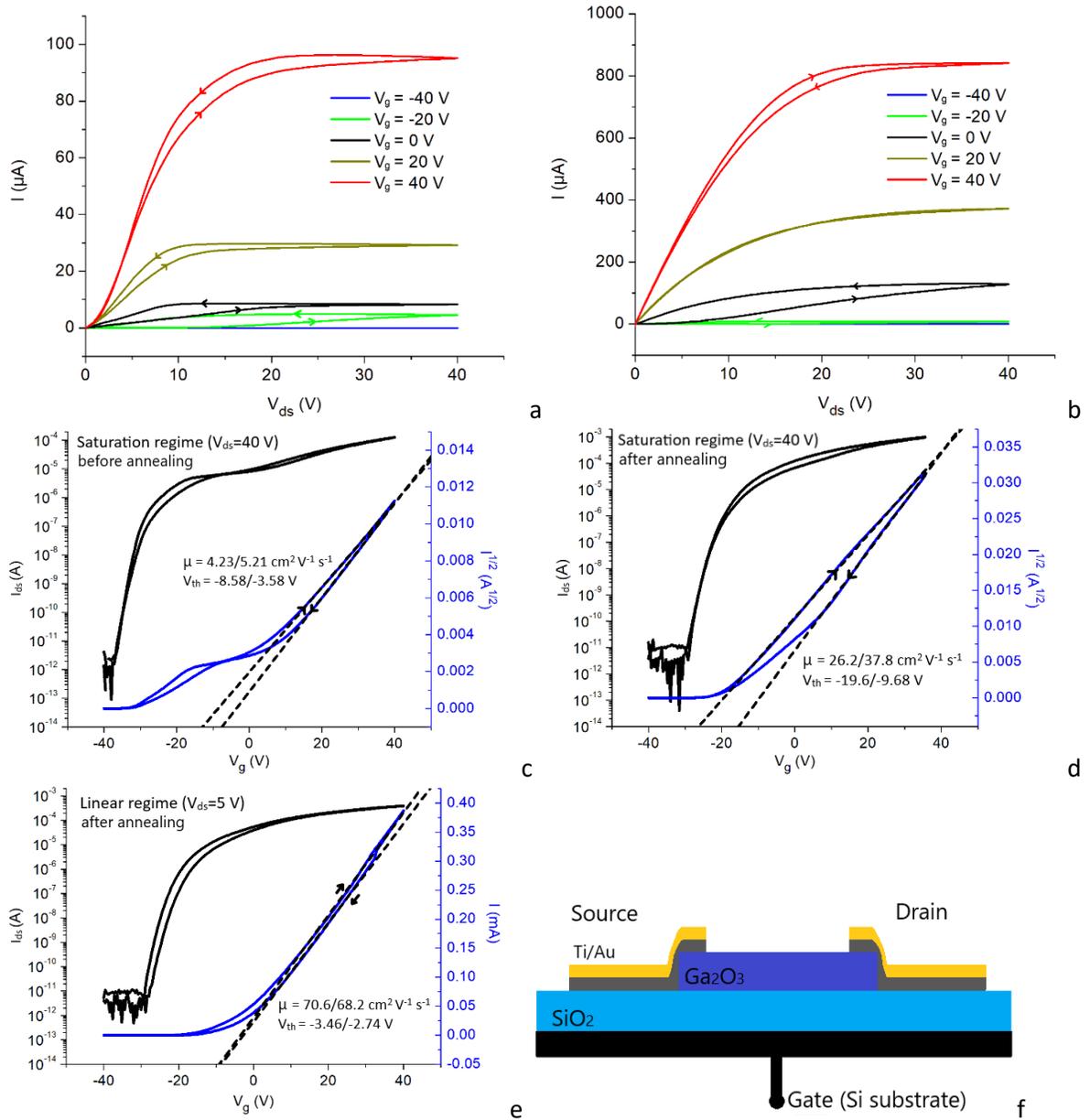

Figure 5 – Output and transfer curves for a representative device. Output curves for 5 different $V_g$: **a** – before annealing, Schottky contacts; **b** – after 500 °C annealing, ohmic contacts. Transfer curves and fits in the saturation regime ($V_{ds}$ = 40 V), shown in logarithmic scale (black, left scale), and $\sqrt{I_{ds}(V_g)}$ curves (blue, right scale). The dashed lines are the linear fits to these curves, from which the mobility and $V_{th}$ are obtained: **c** – before annealing; **d** – after annealing. Transfer curves and fits in the linear regime ($V_{ds}$ = 5 V), shown in logarithmic scale (black, left scale) and in linear scale (blue, right scale). The dashed lines represent the linear fits to the $I_{ds}(V_g)$ curve for which the transconductance and mobility are obtained: **e** – after annealing. Some values obtained from the fits are presented for measurements with increasing/decreasing $V_g$. **f** – Schematic cross-section of the GOOI FET.

Figure 5c shows the transfer curve (in logarithmic scale) for the unannealed device in the saturation regime, for $V_{ds}$ = 40 V, in black; and the $\sqrt{I_{ds}(V_g)}$ curve, in blue. Figure 5d shows the corresponding curves for the device after annealing. Figure 5e shows the transfer curve for the annealed device in the linear regime, for $V_{ds}$= 5 V, in logarithmic scale, in black. However, the blue curve now represents simply the transfer curve in linear scale. The dashed lines represent the fits used to obtain the mobilities ($\mu$) and threshold voltages ($V_{th}$). Details on the data treatment are explained in section 4.2.

Both before and after annealing, it can be seen that for low enough $V_g$ the device can be turned off completely and it is possible to distinguish the off regions of the device, for high negative bias, and the on region, above some voltage threshold. These curves help to better understand the transition between the on and off states. For example, it is possible to see that for the unannealed device, the transition seems to show a hump (Figure 5c), an effect which is absent after annealing (Figure 5d), suggesting an improvement of the device behavior.

Table 1 shows all of the parameters obtained from the transfer curves. Besides $\mu$ and $V_{th}$, values for the subthreshold swing (*SS*), onset voltage ($V_{on}$), on/off ratio ($I_{on}/I_{off}$) and transconductance per unit width ($g_m/w$) are also presented. Table 2 presents values for most of these parameters reported in devices from the literature. In this comparison, it is important to keep in mind that all the literature devices have much lower width and that the membranes in references [9–13] are Sn-doped, which may explain some of the differences.

Before annealing, the obtained value for the mobility is very low, compared with the experimental values found in the literature, since values in $Ga_2O_3$ can reach up to 300 cm$^2$ V$^{-1}$ s$^{-1}$.[105] $I_{ds,max}/w$ is also quite low. The lower values obtained for these parameters were attributed to the presence of Schottky barriers, hindering the mobility of the carriers. In fact, after annealing, the mobility experienced a big increase. Note that this "extrinsic" field-effect mobility does not take into account contact resistance, we attribute the improvement entirely to the change in contact resistance during annealing since we do not expect significant change of $Ga_2O_3$ membrane at this temperature.

| Annealing | No | 500 °C in N$_2$ atmosphere | |
|---|---|---|---|
| Regime | Saturation ($V_{ds}$=40 V) | Linear ($V_{ds}$=5 V) | Saturation ($V_{ds}$=40 V) |
| $\mu$ (cm$^2$ V$^{-1}$ s$^{-1}$) | 4.23/5.21 | 70.6/68.2 | 26.2/37.8 |
| *SS* (mV/dec) | (1.29/1.52)×10$^3$ | (1.33/1.62)×10$^3$ | (1.45/1.60)×10$^3$ |
| $V_{th}$ (V) | −8.58/−3.58 | −3.46/−2.74 | −19.6/−9.68 |
| $I_{on}/I_{off}$ | 2.54×10$^7$ | 7.74×10$^7$ | >2×10$^8$ |
| $V_{on}$ (V) | −36.4 | −29.2/−28.4 | −28.8/−30.4 |
| $g_m/w$ (mS/mm) | —— | 0.084/0.081 | —— |
| $I_{ds,max}/w$ (mA/mm) | 0.91 | 8.0 | |

*Table 1 - Values of the parameters obtained for the FET device, before and after annealing, and in the linear and saturation regime. Since a double sweep was performed, some parameters have values for increasing and for deceasing $V_g$, presented in this order.*

| Ref | $\mu$ (cm$^2$ V$^{-1}$ s$^{-1}$) | SS (mV/dec) | $V_{th}$ (V) | $I_{on}/I_{off}$ | $g_m/w$ (mS/mm) | $I_{ds,max}/w$ (mA/mm) |
|---|---|---|---|---|---|---|
| [9] | 48.8/55.2 | 250/140 | −80/7 | $10^{10}$ | 3.3/4.5 | 600/450 |
| [10] | ----- | 165/150 | −135/2 | $10^{10}$ | 9.2 | 1500/1000 |
| [11] | 21.7/30.2 | 65/65 | −25/−27 | $10^9$ | 13/21 | 325/535 |
| [12] | 82.9 | 80 | −9 | $10^9$ | 35.5 | 580 |
| [13] | 65 | —— | −80 | $6\times10^9$ | —— | 3100 |
| [14] | 184 | 110 | −7.1 | $1.2\times10^9$ | 0.68 | —— |
| [21] | —— | —— | —— | $1.27\times10^7$ | 0.138 | —— |
| [22] | —— | —— | 123 | −23 | $\sim10^8$ | —— | 95.7 |
| [23] | —— | —— | −4.6 | $7.5\times10^6$ | 0.121 | —— |

Table 2 - Values of the parameters found in the literature for FET based on Ga$_2$O$_3$ membranes obtained by mechanical exfoliation. The last three references present phototransistors. For references that include two devices, both values are shown.

The other parameters remained unchanged after annealing for the most part. In particular, *SS* does not change much with annealing and its value is unusually high when compared to the values reported in the literature. High *SS* could indicate a high density of trap states at the interface between the channel and the dielectric, [106] suggesting that these devices could benefit from an interface optimization. Furthermore, the mobility obtained in the linear region is clearly higher than that obtained in saturation, which is not what is usually expected, since more charge traps are filled in saturation. [106] The values obtained in the linear regime may be more accurate since in saturation the higher voltage may lead to threshold voltage shift and instability of the device. [106] $V_{th}$ is found to be negative, meaning that the device works in depletion-mode. [107] For this device, $V_{th}$ is not far from zero, which is important for low power consumption. [106] In the literature, it is shown that $V_{th}$ may depend on the membrane thickness, [9] so, using ion-beam-assisted exfoliation, it should be possible to control the thickness of the membranes by simply changing the ion energy, [31] and thus also the threshold voltage of the respective device, which is something to be explored. Finally, $I_{on}/I_{off}$ was found to reach $2\times10^8$ which was the maximum possible value that could be measured under the experimental conditions, considering noise under 5 pA and setting a compliance at 1 mA.

## 3. Conclusions

In this work, MSM structures based on ion-beam-assisted exfoliation of β-Ga$_2$O$_3$ membranes were fabricated and tested as MSM photodetectors and GOOI FET. Both types of devices displayed a rather good behavior, indicating the viability of employing membranes obtained from this novel technique in high-performance devices. Ti/Au contacts to the membrane were used, and these were found to be rectifying if no annealing is performed, but reaching their optimal ohmic behavior after annealing at 500 °C for 1 min in N$_2$ atmosphere. One of the MSM photodetectors discussed here displayed an already high responsivity of $2.6\times10^4$ A/W, with a detectivity of $2.4\times10^{14}$ Jones and a *PDCR*= $3.0\times10^2$. Using a microprobe, it was possible to distinguish response mechanisms happening at the contacts from mechanisms related to the conductive channel along the membrane. There is a lot of room for optimization on these devices, as there is still a lot of variability between devices. A different device with very low dark current was measured to have *R*= $3.3\times10^3$ A/W but a lot higher *D\**= $2.6\times10^{16}$ Jones and *PDCR*= $2.6\times10^7$. A conclusion that can be taken out of this is that having control over the dark current is an important next step in the improvement of these devices. A very interesting way of achieving this is by developing phototransistors, so that a gate voltage can turn off the device in the dark, sharply decreasing the dark current. This has found much success in the literature, [22,29] as seen in Figure 3c, and may also prove successful here, given that our GOOI FET representative device shows a high $I_{on}/I_{off}$, reaching at least $2\times10^8$, and a $V_{th}$ not far from zero. Despite these promising results, some device performance parameters, namely the strong PPC in photodetectors and the high $SS$ in FET remain underwhelming compared to the best results found in the literature, pointing to a high density of defect levels. Further work is necessary to pinpoint the type and origin of these defects which can arise due to the implantation process itself or due to the rolling and unrolling of microtubes, which is likely to introduce extended defects such as stacking faults or dislocations into the lattice.

## 4. Experimental Section and Data Treatment

### 4.1. Experimental Details

In this study, the unintentionally doped (UID) bulk β-$Ga_2O_3$ crystals with (100) surface that were used were purchased from Novel Crystal Technology, Inc.. Ion-beam-assisted exfoliation was performed by implantation of 250 keV Cr ions. This was done at the 210 kV high flux ion implanter of the Laboratory of Accelerators of Instituto Superior Técnico (IST), Universidade de Lisboa, [108] using $Cr^{2+}$ ions with a fluence of $5\times10^{14}$ cm$^{-2}$ and a flux ≤$1\times10^{12}$ cm$^{-2}$s$^{-1}$ at an angle of 7°; and at the 500 kV implanter of the Ion Beam Centre of the Helmholtz Zentrum Dresden-Rossendorf (IBC-HZDR) using $Cr^+$ ions with a fluence of $1\times10^{14}$ cm$^{-2}$ and a flux of $1.5\times10^{12}$ cm$^{-2}$s$^{-1}$. A schematic of the exfoliation process is shown in Figure 6a. Optical microscopy inspection showed the formation of a good amount of tubes on every sample.

In order to transfer the tubes to the intended substrates, a pick and place technique was employed, using the set-up presented in Figure 6b, in which the tubes are manipulated by tweezers, controlled using micropositioners. A large number of tubes was moved to $Al_2O_3$ substrates, cut from a (650±20) µm thick c-plane substrate purchased from Siegert Wafer, as well as to p-type Si/$SiO_2$ substrates, from the same manufacturer. The p-Si has a thickness of (279±25) µm, a resistivity <0.005 Ω cm, and a coating of 290 nm ±5% dry thermal oxide. After transfer, as shown in Figure 6c, the tubes were unrolled by thermal annealing, using an AS-One system from AnnealSys at 1000 °C for 60 s in $N_2$ atmosphere (at 1 atm) and a heating ramp of 2 °C/s. Annealing at this temperature yields an efficient removal of implantation damage and membranes of bulk-like crystalline quality. [31]

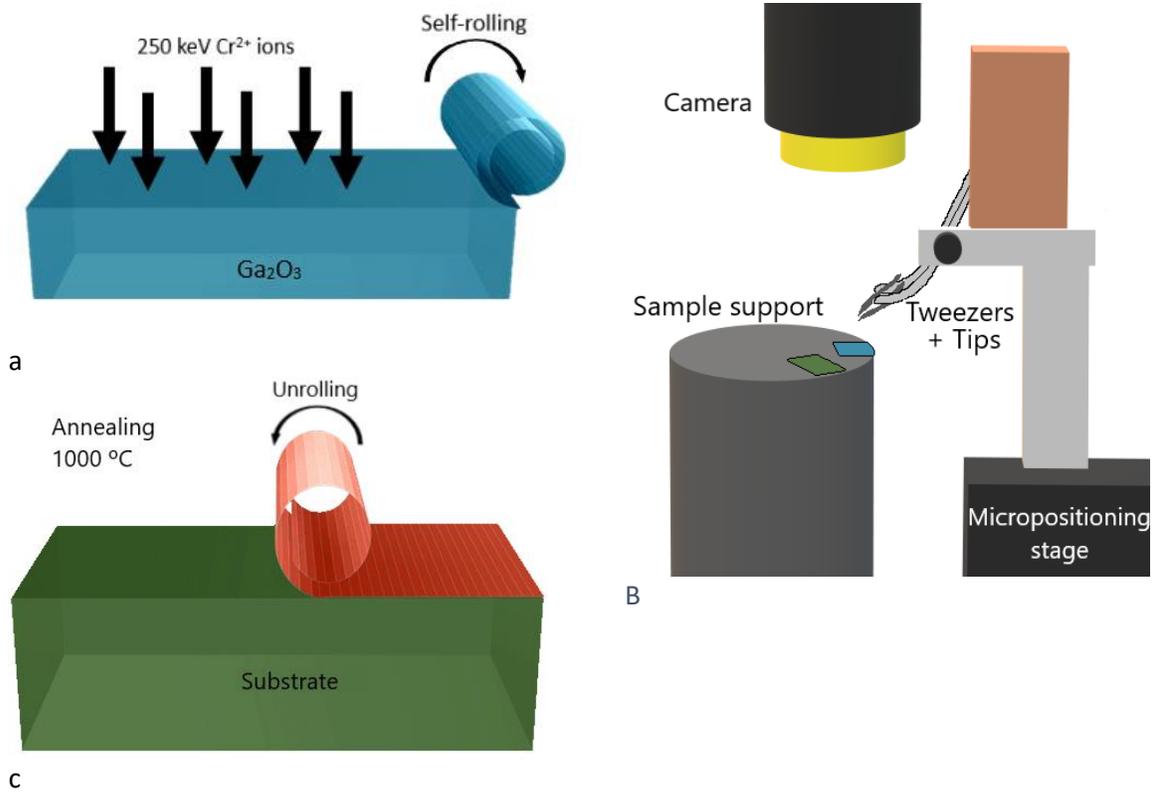

*Figure 6 – **a** – Schematics of the ion-beam assisted exfoliation process, producing tubes; [31] **b** – Schematics of the set-up used for the pick and placed technique that allowed transfer of tubes from the bulk crystal surface to the desired substrate. The crystal and the new substrate lie on a support that can be rotated. A pair of probe tips are attached to tweezers, which manipulate the tubes. The process is observed by a camera placed above the samples and controlled by micropositioners; **c** – Unrolling of the tubes into membranes upon annealing.*

The next step is to create the contacts by depositing Ti/Au on both ends of each membrane. For the Au layer, 100 nm were always used; for Ti, thicknesses of 100 nm and 300 nm were used, without any significant difference in behavior being observed. TEM grids (grids usually used for transmission electron microscopy) with a diameter of 3.05 mm and a thickness of 25 µm were used as shadow masks for the deposition, with 50 mesh grids (425×425 µm$^2$ hole and 83 µm bar) and 100 mesh grids (204×204 µm$^2$ hole and 50 µm bar) being used according to the membranes' size. The deposition was performed using an Alcatel SCM 450 sputtering tool, using DC (direct current, 40 W) sputtering for Ti and RF (radio frequency, 20 W) sputtering for Au. In both cases, the working pressure was close to 3.5 mTorr, with an Ar flux of 20.0 sccm. This results in an MSM structure (Figure 1a), which already allows to test the contacts, and also concludes the devices intended to be tested as MSM photodetectors. Photodetectors were fabricated on $Al_2O_3$ substrates because $Al_2O_3$ is transparent allowing light excitation in the contact region and thus mitigating the shadowing effect of the contacts. To test devices as FET, they are fabricated on $Si/SiO_2$ substrates, and a region of the $SiO_2$ layer is manually removed using a diamond-tipped pen, allowing a voltage to be imposed on the Si itself, which becomes the gate of the FET.

For the characterization, the current-voltage (*I-V*) and current-time characteristics of the devices were obtained using a Keysight Technologies B1500A Semiconductor Device Analyzer and a homemade system consisting of a CPX200DP AIM-TTI voltage source, a Keithley 6485 picoammeter, a Keithley 6221 current source and a Keithley 2182A nanovoltmeter, where its control and data acquisition were performed through a PC with LabVIEW software.

To optimize the contacts, MSM structures were subjected to an isochronal annealing study at temperatures ranging from 200 to 800 °C. This process was done in an AS-One rapid thermal processing system in a $N_2$ atmosphere and using the same heating ramp (2 °C/s) and holding time (60 s) as used for the membrane fabrication.

For photodetector characterization, a ThorLabs M365D2 LED, emitting at a peak wavelength of (365±5) nm, and a Roithner LaserTechnik UVTOP240 LED, emitting at a peak wavelength of (245±5) nm, were used, with their power calibrated using a ThorLabs FDS010 photodiode. LED245 had a power density of 4.0 W/m$^2$ and LED365 a power density of 4.7 W/m$^2$, measured with the photodiode at the same position as the samples. Light was made to reach the devices from underneath the substrate. LED245 was directly placed at less than 1 cm below the sample, while the LED365 light was guided through an optical fiber cable, placed at around 1 mm below the sample.

Using a 1.6 MeV He$^+$ beam for Rutherford backscattering spectrometry (RBS) on 15 membranes, a thickness of (306±79) nm was obtained. This was done using a 2.5 MV Van de Graaff accelerator and an Oxford Microbeams microprobe setup with a quadrupole triplet for focusing. The surface barrier detector was directed at an angle of 140° in a Cornell geometry. Using the same µ-probe system, the I-V response of some devices was measured as a 2.0 MeV H$^+$ beam was directed at specific regions of the devices. The focused beam has dimensions of 3×4 µm$^2$ and used to sweep areas of around 200 to 250 µm$^2$ on the devices. The simulation of the interaction of the beam with the materials that compose the device employed stopping power values were given by the SRIM software (Stopping and Range of Ions in Matter). [102]

### 4.2. Data treatment

Three important parameters to evaluate photodetectors are the responsivity (*R*), photo-to-dark current ratio (*PDCR*) and detectivity (*D\**), which are defined as $R = \frac{I_{\text{photo}} - I_{\text{dark}}}{SP_{\text{light}}}$, $PDCR = \frac{I_{\text{photo}} - I_{\text{dark}}}{I_{\text{dark}}}$ and $D* = \frac{R\sqrt{S}}{\sqrt{2qI_{\text{dark}}}}$, [3] where $I_{\text{photo}}$ is the photocurrent, $I_{\text{dark}}$ is the dark current, $P_{\text{light}}$ is the power density of the incident light, *S* is the active area of the photodetector (taken as the full area of the membrane, in this case) and *q* is the elementary charge. Responsivity relates to photoconductive gain according to the relation *R* = *ηg*(*λq*)/(*hc*), [109] where *g* is the photoelectric current gain, *η* is the internal quantum efficiency, *λ* is the wavelength, *h* is Planck's constant and *c* is the speed of light in vaccum. The product *ηg* is often used to characterize photodetectors, and it is called the external quantum efficiency (*EQE*). [66]

In order to characterize FETs, transfer curves, of the measured source-drain current ($I_{\text{ds}}$) as a function of the gate voltage ($V_{\text{g}}$), are very useful. Using these, parameters can be obtained from the linear and from the saturation region by a set of formulas, [106,110] beginning from the behavior of a FET in the linear and saturation regions we get respectively:

$$I_{\text{ds}} = \frac{w\epsilon_{\text{d}}}{Ld} \mu_{\text{lin}}(V_{\text{g}} - V_{\text{th}})V_{\text{ds}} \quad \text{and} \quad I_{\text{ds}} = \frac{w\epsilon_{\text{d}}}{2Ld} \mu_{\text{sat}}(V_{\text{g}} - V_{\text{th}})^2 \quad ,$$

where *w* is the channel width, *L* is the channel length, *d* is dielectric thickness and $\varepsilon_d$ is the dielectric permittivity. The threshold voltage ($V_{th}$), mobility ($\mu$), subthreshold swing (*SS*), onset voltage ($V_{on}$) and on/off ratio ($I_{on}/I_{off}$) can be obtained using data from both regions.

In the linear regime, several parameters can be obtained from linear fit to the transfer curve, starting with the transconductance $g_m = \frac{\partial I_{ds}}{\partial V_g}$, corresponding to the slope of the curve, which is used to calculate the mobility $\mu_{lin} = \frac{g_m L d}{\epsilon_d w V_{ds}}$. $V_{th}$ is found by extrapolating the linear fit to zero. *SS* can be obtained by a linear fit to the transfer curves in logarithmic scale, since $SS = \frac{\partial V_g}{\partial \log_{10}(I_{ds})}$.

In the saturation regime, $\mu_{sat}$ and $V_{th}$ are obtained from linear fits to the $\sqrt{I_{ds}(V_g)}$ curve. $\mu_{sat}$ is related to the slope ($\mu_{sat} = \frac{2Ld}{\epsilon_d w}\left(\frac{\partial \sqrt{I_{ds}}}{\partial V_g}\right)^2$) and $V_{th}$ is again found by extrapolating the linear fit to zero.

# Acknowledgments

The authors acknowledge the financial support from the Portuguese Foundation for Science and Technology (FCT) via the IonProGO project (2022.05329.PTDC, http://doi.org/10.54499/2022.05329.PTDC), via the INESC MN Research Unit funding (UID/05367/2020) through Pluriannual BASE and PROGRAMATICO financing, and via the C2TN funding (UIDB/04349/2020). This work has also received funding from the National funds through FCT under the PhD grant 2022.09585.BD. The implantations were performed under proposal 26001 of the ReMade@ARI project (https://doi.org/10.3030/101058414), funded by the European Union as part of the Horizon Europe call HORIZON-INFRA-2021-SERV-01 under grant agreement number 101058414 and co-funded by UK Research and Innovation (UKRI) under the UK government's Horizon Europe funding guarantee (grant number 10039728) and by the Swiss State Secretariat for Education, Research and Innovation (SERI) under contract number 22.00187. Views and opinions expressed are however those of the author(s) only and do not necessarily reflect those of the European Union or the UK Science and Technology Facilities Council or the Swiss State Secretariat for Education, Research and Innovation (SERI). Neither the European Union nor the granting authorities can be held responsible for them.

# Bibliography

[1] C. Lu, X. Ji, Z. Liu, X. Yan, N. Lu, P. Li, W. Tang, *J. Phys. Appl. Phys.* **2022**, *55*, 463002.
[2] L. A. M. Lyle, *J. Vac. Sci. Technol. A* **2022**, *40*, 060802.
[3] Z. Liu, W. Tang, *J. Phys. Appl. Phys.* **2023**, *56*, 093002.
[4] D. Kaur, R. Dahiya, M. Kumar, *J. Vac. Sci. Technol. A* **2023**, *41*, 043410.
[5] M. Higashiwaki, G. H. Jessen, *Appl. Phys. Lett.* **2018**, *112*, 060401.
[6] S. J. Pearton, J. Yang, P. H. Cary, F. Ren, J. Kim, M. J. Tadjer, M. A. Mastro, *Appl. Phys. Rev.* **2018**, *5*, 011301.
[7] Y. Tomm, P. Reiche, D. Klimm, T. Fukuda, *J. Cryst. Growth* **2000**, *220*, 510.
[8] W. S. Hwang, A. Verma, H. Peelaers, V. Protasenko, S. Rouvimov, H. (Grace) Xing, A. Seabaugh, W. Haensch, C. V. De Walle, Z. Galazka, M. Albrecht, R. Fornari, D. Jena, *Appl. Phys. Lett.* **2014**, *104*, 203111.

[9] H. Zhou, M. Si, S. Alghamdi, G. Qiu, L. Yang, P. D. Ye, *IEEE Electron Device Lett.* **2017**, *38*, 103.
[10] H. Zhou, K. Maize, G. Qiu, A. Shakouri, P. D. Ye, *Appl. Phys. Lett.* **2017**, *111*, 092102.
[11] H. Zhou, K. Maize, J. Noh, A. Shakouri, P. D. Ye, *ACS Omega* **2017**, *2*, 7723.
[12] D. Lei, K. Han, Y. Wu, Z. Liu, X. Gong, *IEEE J. Electron Devices Soc.* **2019**, *7*, 596.
[13] Z. Li, Y. Liu, A. Zhang, Q. Liu, C. Shen, F. Wu, C. Xu, M. Chen, H. Fu, C. Zhou, *Nano Res.* **2019**, *12*, 143.
[14] J. Ma, H. J. Cho, J. Heo, S. Kim, G. Yoo, *Adv. Electron. Mater.* **2019**, *5*, 1800938.
[15] S. Oh, M. A. Mastro, M. J. Tadjer, J. Kim, *ECS J. Solid State Sci. Technol.* **2017**, *6*, Q79.
[16] H. Kim, S. Tarelkin, A. Polyakov, S. Troschiev, S. Nosukhin, M. Kuznetsov, J. Kim, *ECS J. Solid State Sci. Technol.* **2020**, *9*, 045004.
[17] Z. Li, Y. Cheng, Y. Xu, Z. Hu, W. Zhu, D. Chen, Q. Feng, H. Zhou, J. Zhang, C. Zhang, Y. Hao, *IEEE Electron Device Lett.* **2020**, *41*, 1794.
[18] S. Oh, C.-K. Kim, J. Kim, *ACS Photonics* **2018**, *5*, 1123.
[19] G. Zeng, X.-X. Li, Y.-C. Li, D.-B. Chen, Y.-C. Chen, X.-F. Zhao, N. Chen, T.-Y. Wang, D. W. Zhang, H.-L. Lu, *ACS Appl. Mater. Interfaces* **2022**, *14*, 16846.
[20] N. Alwadai, Z. Alharbi, F. Alreshidi, S. Mitra, B. Xin, H. Alamoudi, K. Upadhyaya, M. N. Hedhili, I. S. Roqan, *ACS Appl. Mater. Interfaces* **2023**, *15*, 12127.
[21] S. Oh, J. Kim, F. Ren, S. J. Pearton, J. Kim, *J. Mater. Chem. C* **2016**, *4*, 9245.
[22] Z. Li, Z. Feng, Y. Xu, Q. Feng, W. Zhu, D. Chen, H. Zhou, J. Zhang, C. Zhang, Y. Hao, *IEEE Electron Device Lett.* **2021**, *42*, 545.
[23] S. Kim, S. Oh, J. Kim, *ACS Photonics* **2019**, *6*, 1026.
[24] Y. Liu, L. Du, G. Liang, W. Mu, Z. Jia, M. Xu, Q. Xin, X. Tao, A. Song, *IEEE Electron Device Lett.* **2018**, *39*, 1696.
[25] S. Yu, X. Zhao, M. Ding, P. Tan, X. Hou, Z. Zhang, W. Mu, Z. Jia, X. Tao, G. Xu, S. Long, *IEEE Electron Device Lett.* **2021**, *42*, 383.
[26] W. Liu, M. Peng, Y. Yu, Z. Zheng, P. Jian, Y. Zhao, Y. Zeng, D. Xu, M. Chen, Y. Luo, C. Chen, J. Dai, F. Wu, *Adv. Opt. Mater.* **2024**, *12*, 2302746.
[27] J. Ahn, J. Ma, D. Lee, Q. Lin, Y. Park, O. Lee, S. Sim, K. Lee, G. Yoo, J. Heo, *ACS Photonics* **2021**, *8*, 557.
[28] S. Kim, J. Kim, *Appl. Phys. Lett.* **2020**, *117*, 261101.
[29] Z. Li, Z. Feng, Y. Huang, Y. Xu, Z. Zhang, Q. Feng, W. Zhu, D. Chen, H. Zhou, J. Zhang, C. Zhang, Y. Hao, *IEEE Trans. Electron Devices* **2022**, *69*, 3807.
[30] K. Lorenz, M. Peres, E. Alves, J. Rocha, *Process of Production of Roll and Submicrometric Membrane of $Ga_2O_3$ by Ion Implantation*, **2022**, PT 117063; EU PCT/PT/2022/050006; USA R6157.0189/189US.
[31] D. M. Esteves, R. He, C. Bazioti, S. Magalhães, M. C. Sequeira, L. F. Santos, A. Azarov, A. Kuznetsov, F. Djurabekova, K. Lorenz, M. Peres, **2025**, DOI 10.48550/arXiv.2501.13055.
[32] Y. Yao, R. F. Davis, L. M. Porter, *J. Electron. Mater.* **2017**, *46*, 2053.
[33] M.-H. Lee, R. L. Peterson, *ECS J. Solid State Sci. Technol.* **2019**, *8*, Q3176.
[34] M.-H. Lee, R. L. Peterson, *APL Mater.* **2019**, *7*, 022524.
[35] M.-H. Lee, T.-S. Chou, S. Bin Anooz, Z. Galazka, A. Popp, R. L. Peterson, *APL Mater.* **2022**, *10*, 091105.
[36] Z. Y. Zhang, C. H. Jin, X. L. Liang, Q. Chen, L.-M. Peng, *Appl. Phys. Lett.* **2006**, *88*, 073102.
[37] E. H. Rhoderick, R. H. Williams, *Metal-Semiconductor Contacts*, Clarendon Press, **1988**.
[38] Y. Liu, Z. Y. Zhang, Y. F. Hu, C. H. Jin, L.-M. Peng, *J. Nanosci. Nanotechnol.* **2008**, *8*, 252.
[39] S. M. Sze, K. K. Ng, *Physics of Semiconductor Devices*, Wiley-Interscience, Hoboken, N.J, **2007**.

[40] B. Qiao, Z. Zhang, X. Xie, B. Li, K. Li, X. Chen, H. Zhao, K. Liu, L. Liu, D. Shen, *J. Phys. Chem. C* **2019**, *123*, 18516.
[41] D. Y. Guo, Z. P. Wu, Y. H. An, X. C. Guo, X. L. Chu, C. L. Sun, L. H. Li, P. G. Li, W. H. Tang, *Appl. Phys. Lett.* **2014**, *105*, 023507.
[42] S. M. Sze, M.-K. Lee, M. K. Lee, *Semiconductor Devices, Physics and Technology*, Wiley, Hoboken, N.J, **2012**.
[43] S. M. Sze, D. J. Coleman, A. Loya, *Solid-State Electron.* **1971**, *14*, 1209.
[44] Y. Qin, L. Li, Z. Yu, F. Wu, D. Dong, W. Guo, Z. Zhang, J. Yuan, K. Xue, X. Miao, S. Long, *Adv. Sci.* **2021**, *8*, 2101106.
[45] Y. Long, K. Ba, J. Liu, X. Deng, Y. Di, K. Xiong, Y. Chen, X. Wang, C. Liu, Z. Li, D. Liu, X. Fang, Q. Liu, J. Wang, *Mater. Today Electron.* **2024**, *10*, 100116.
[46] L.-X. Qian, Z.-H. Wu, Y.-Y. Zhang, P. T. Lai, X.-Z. Liu, Y.-R. Li, *ACS Photonics* **2017**, *4*, 2203.
[47] G. C. Hu, C. X. Shan, N. Zhang, M. M. Jiang, S. P. Wang, D. Z. Shen, *Opt. Express* **2015**, *23*, 13554.
[48] Y. Xu, Y. Cheng, Z. Li, D. Chen, S. Xu, Q. Feng, W. Zhu, Y. Zhang, J. Zhang, C. Zhang, Y. Hao, *Adv. Mater. Technol.* **2021**, *6*, 2001296.
[49] C. Xie, X.-T. Lu, M.-R. Ma, X.-W. Tong, Z.-X. Zhang, L. Wang, C.-Y. Wu, W.-H. Yang, L.-B. Luo, *Adv. Opt. Mater.* **2019**, *7*, 1901257.
[50] L.-X. Qian, H.-F. Zhang, P. T. Lai, Z.-H. Wu, X.-Z. Liu, *Opt. Mater. Express* **2017**, *7*, 3643.
[51] Z. X. Jiang, Z. Y. Wu, C. C. Ma, J. N. Deng, H. Zhang, Y. Xu, J. D. Ye, Z. L. Fang, G. Q. Zhang, J. Y. Kang, T.-Y. Zhang, *Mater. Today Phys.* **2020**, *14*, 100226.
[52] G. Shen, Z. Liu, K. Tang, S. Sha, L. Li, C.-K. Tan, Y. Guo, W. Tang, *Sci. China Technol. Sci.* **2023**, *66*, 3259.
[53] Z. Zheng, B. Qiao, Z. Zhang, X. Huang, X. Xie, B. Li, X. Chen, K. Liu, L. Liu, D. Shen, *IEEE Trans. Electron Devices* **2022**, *69*, 4362.
[54] X. Gao, T. Xie, J. Wu, J. Fu, X. Gao, M. Xie, H. Zhao, Y. Wang, Z. Shi, *Appl. Phys. Lett.* **2024**, *125*, 172103.
[55] Q. Zhang, D. Dong, T. Zhang, T. Zhou, Y. Yang, Y. Tang, J. Shen, T. Wang, T. Bian, F. Zhang, W. Luo, Y. Zhang, Z. Wu, *ACS Nano* **2023**, *17*, 24033.
[56] W. Zhang, W. Wang, J. Zhang, T. Zhang, L. Chen, L. Wang, Y. Zhang, Y. Cao, L. Ji, J. Ye, *ACS Appl. Mater. Interfaces* **2023**, *15*, 10868.
[57] Q. Wan, A. Zhang, W. Cao, Y. Bai, B. Wang, H. Cheng, G. Wang, *Opt. Express* **2024**, *32*, 32322.
[58] X. Zhu, Y. Wu, G. Li, W. Lu, *J. Alloys Compd.* **2023**, *953*, 170109.
[59] P. Yaghoubizadeh, M. J. Eshraghi, S. Hajati, N. Naderi, *ACS Appl. Electron. Mater.* **2023**, *5*, 4220.
[60] S. Fu, Y. Wang, Y. Han, B. Li, Y. Zhang, J. Ma, Z. Fu, H. Xu, Y. Liu, *IEEE Electron Device Lett.* **2022**, *43*, 1511.
[61] H. Wang, J. Ma, L. Cong, D. Song, L. Fei, P. Li, B. Li, Y. Liu, *Mater. Today Phys.* **2022**, *24*, 100673.
[62] L. Bao, Z. Liang, S. Kuang, B. Xiao, K. H. L. Zhang, X. Xu, Q. Cheng, *J. Mater. Chem. C* **2024**, *12*, 14876.
[63] H. Chen, P. Lv, K. Peng, P. Li, N. Ji, L. Wang, D. Xue, C. Chen, *Vacuum* **2024**, *226*, 113335.
[64] Y. Sui, H. Liang, W. Huo, X. Zhan, T. Zhu, Z. Mei, *Mater. Futur.* **2024**, *3*, 015701.
[65] X. Li, F. Xu, X. Wang, J. Luo, K. Ding, L. Ye, H. Li, Y. Xiong, P. Yu, C. Kong, L. Ye, H. Zhang, W. Li, *Phys. Status Solidi RRL – Rapid Res. Lett.* **2024**, *18*, 2200512.
[66] X. Ji, X. Yin, Y. Yuan, S. Yan, X. Li, Z. Ding, X. Zhou, J. Zhang, Q. Xin, A. Song, *J. Alloys Compd.* **2023**, *933*, 167735.

[67] Z. Liu, X. Wang, Y. Liu, D. Guo, S. Li, Z. Yan, C.-K. Tan, W. Li, P. Li, W. Tang, *J. Mater. Chem. C* **2019**, *7*, 13920.
[68] Z. Liu, S. Li, Z. Yan, Y. Liu, Y. Zhi, X. Wang, Z. Wu, P. Li, W. Tang, *J. Mater. Chem. C* **2020**, *8*, 5071.
[69] W.-Y. Jiang, Z. Liu, S. Li, Z.-Y. Yan, C.-L. Lu, P.-G. Li, Y.-F. Guo, W.-H. Tang, *IEEE Sens. J.* **2021**, *21*, 18663.
[70] Z. Peng, X. Hou, Z. Han, Z. Gan, C. Li, F. Wu, S. Bai, S. Yu, Y. Liu, K. Yang, X. Feng, H. Zhan, X. Zhao, G. Xu, S. Long, *Adv. Funct. Mater.* **2024**, 2405277.
[71] Y. Qin, H. Dong, S. Long, Q. He, G. Jian, Y. Zhang, X. Zhou, Y. Yu, X. Hou, P. Tan, Z. Zhang, Q. Liu, H. Lv, M. Liu, *IEEE Electron Device Lett.* **2019**, *40*, 742.
[72] Y. Qin, S. Long, Q. He, H. Dong, G. Jian, Y. Zhang, X. Hou, P. Tan, Z. Zhang, Y. Lu, C. Shan, J. Wang, W. Hu, H. Lv, Q. Liu, M. Liu, *Adv. Electron. Mater.* **2019**, *5*, 1900389.
[73] Z. Han, H. Liang, W. Huo, X. Zhu, X. Du, Z. Mei, *Adv. Opt. Mater.* **2020**, *8*, 1901833.
[74] X.-X. Li, G. Zeng, Y.-C. Li, H. Zhang, Z.-G. Ji, Y.-G. Yang, M. Luo, W.-D. Hu, D. W. Zhang, H.-L. Lu, *Npj Flex. Electron.* **2022**, *6*, 47.
[75] T. Chen, J. Zhang, X. Zhang, C. Chen, L. Zhang, Y. Hu, Y. Ma, X. Wei, X. Zhou, W. Tang, A. Yang, B. Li, S. Dai, L. Xu, W. Shi, H. Fu, Y. Fan, Y. Cai, Z. Zeng, K. Zhang, B. Zhang, *IEEE Sens. J.* **2023**, *23*, 15504.
[76] X. Ji, Y. Yuan, X. Yin, S. Yan, Z. Ding, J. Zhang, Q. Xin, A. Song, *IEEE Electron Device Lett.* **2023**, *44*, 436.
[77] R. Lin, W. Zheng, D. Zhang, Z. Zhang, Q. Liao, L. Yang, F. Huang, *ACS Appl. Mater. Interfaces* **2018**, *10*, 22419.
[78] W.-Y. Kong, G.-A. Wu, K.-Y. Wang, T.-F. Zhang, Y.-F. Zou, D.-D. Wang, L.-B. Luo, *Adv. Mater.* **2016**, *28*, 10725.
[79] B. Zhao, F. Wang, H. Chen, Y. Wang, M. Jiang, X. Fang, D. Zhao, *Nano Lett.* **2015**, *15*, 3988.
[80] S. Qi, J. Liu, J. Yue, X. Ji, J. Shen, Y. Yang, J. Wang, S. Li, Z. Wu, W. Tang, *J. Mater. Chem. C* **2023**, *11*, 8454.
[81] Q. Zhang, N. Li, T. Zhang, D. Dong, Y. Yang, Y. Wang, Z. Dong, J. Shen, T. Zhou, Y. Liang, W. Tang, Z. Wu, Y. Zhang, J. Hao, *Nat. Commun.* **2023**, *14*, 418.
[82] W. E. Mahmoud, *Sol. Energy Mater. Sol. Cells* **2016**, *152*, 65.
[83] Y. Han, Y. Wang, S. Fu, J. Ma, H. Xu, B. Li, Y. Liu, *Small* **2023**, *19*, 2206664.
[84] Y. Wang, Z. Lin, J. Ma, Y. Wu, H. Yuan, D. Cui, M. Kang, X. Guo, J. Su, J. Miao, Z. Shi, T. Li, J. Zhang, Y. Hao, J. Chang, *InfoMat* **2024**, *6*, e12503.
[85] U. Varshney, A. Sharma, L. Goswami, J. Tawale, G. Gupta, *Vacuum* **2023**, *217*, 112570.
[86] G. Zeng, M.-R. Zhang, Y.-C. Chen, X.-X. Li, D.-B. Chen, C.-Y. Shi, X.-F. Zhao, N. Chen, T.-Y. Wang, D. W. Zhang, H.-L. Lu, *Mater. Today Phys.* **2023**, *33*, 101042.
[87] V. Janardhanam, J.-H. Kim, I. Jyothi, H.-H. Jung, S.-J. Kim, K.-H. Shim, C.-J. Choi, *Colloids Surf. Physicochem. Eng. Asp.* **2024**, *693*, 134079.
[88] A. Y. Polyakov, E. B. Yakimov, I. V. Shchemerov, A. A. Vasilev, A. I. Kochkova, V. I. Nikolaev, S. J. Pearton, *J. Phys. Appl. Phys.* **2024**, *58*, 063002.
[89] E. B. Yakimov, A. Y. Polyakov, I. V. Shchemerov, N. B. Smirnov, A. A. Vasilev, A. I. Kochkova, P. S. Vergeles, E. E. Yakimov, A. V. Chernykh, M. Xian, F. Ren, S. J. Pearton, *J. Alloys Compd.* **2021**, *879*, 160394.
[90] Y. Xu, X. Chen, D. Zhou, F. Ren, J. Zhou, S. Bai, H. Lu, S. Gu, R. Zhang, Y. Zheng, J. Ye, *IEEE Trans. Electron Devices* **2019**, *66*, 2276.
[91] P. Mukhopadhyay, I. Hatipoglu, Y. K. Frodason, J. B. Varley, M. S. Williams, D. A. Hunter, N. K. Gunasekar, P. R. Edwards, R. W. Martin, F. Wu, A. Mauze, J. S. Speck, W. V. Schoenfeld, *Appl. Phys. Lett.* **2022**, *121*, 111105.

[92] C. Borelli, A. Bosio, A. Parisini, M. Pavesi, S. Vantaggio, R. Fornari, *Mater. Sci. Eng. B* **2022**, *286*, 116056.
[93] Y. Dan, X. Zhao, K. Chen, A. Mesli, *ACS Photonics* **2018**, *5*, 4111.
[94] J. P. Vilcot, J. L. Vaterkowski, D. Decoster, M. Constant, *Electron. Lett.* **1984**, *20*, 86.
[95] J. A. Garrido, E. Monroy, I. Izpura, E. Muñoz, *Semicond. Sci. Technol.* **1998**, *13*, 563.
[96] A. Asteriti, G. Verzellesi, G. Sozzi, A. Baraldi, P. Mazzolini, A. Moumen, A. Parisini, M. Pavesi, M. Bosi, R. Mosca, L. Seravalli, R. Fornari, *Phys. Status Solidi B* **2025**, *n/a*, 2400581.
[97] D. Cammi, C. Ronning, *Adv. Condens. Matter Phys.* **2014**, *2014*, 184120.
[98] D. R. Pereira, M. Peres, L. C. Alves, J. G. Correia, C. Díaz-Guerra, A. G. Silva, E. Alves, K. Lorenz, *Surf. Coat. Technol.* **2018**, *355*, 50.
[99] T. C. Lovejoy, R. Chen, X. Zheng, E. G. Villora, K. Shimamura, H. Yoshikawa, Y. Yamashita, S. Ueda, K. Kobayashi, S. T. Dunham, F. S. Ohuchi, M. A. Olmstead, *Appl. Phys. Lett.* **2012**, *100*, 181602.
[100] H. Zhou, L. Cong, J. Ma, B. Li, H. Xu, Y. Liu, *Chin. Phys. B* **2021**, *30*, 126104.
[101] S. Jeon, S.-E. Ahn, I. Song, C. J. Kim, U.-I. Chung, E. Lee, I. Yoo, A. Nathan, S. Lee, K. Ghaffarzadeh, J. Robertson, K. Kim, *Nat. Mater.* **2012**, *11*, 301.
[102] J. F. Ziegler, J. P. Biersack, in *Treatise Heavy-Ion Sci. Vol. 6 Astrophys. Chem. Condens. Matter* (Ed.: D. A. Bromley), Springer US, Boston, MA, **1985**, pp. 93–129.
[103] M. E. Ingebrigtsen, A. Yu. Kuznetsov, B. G. Svensson, G. Alfieri, A. Mihaila, U. Badstübner, A. Perron, L. Vines, J. B. Varley, *APL Mater.* **2018**, *7*, 022510.
[104] E. B. Yakimov, A. Y. Polyakov, N. B. Smirnov, I. V. Shchemerov, J. Yang, F. Ren, G. Yang, J. Kim, S. J. Pearton, *J. Appl. Phys.* **2018**, *123*, 185704.
[105] M. Higashiwaki, K. Sasaki, A. Kuramata, T. Masui, S. Yamakoshi, *Appl. Phys. Lett.* **2012**, *100*, 013504.
[106] W.-Y. Lee, J. Mei, Z. Bao, in *Mater. Energy*, World Scientific, **2016**, pp. 19–83.
[107] E. Fortunato, P. Barquinha, R. Martins, *Adv. Mater.* **2012**, *24*, 2945.
[108] E. Alves, K. Lorenz, N. Catarino, M. Peres, M. Dias, R. Mateus, L. C. Alves, V. Corregidor, N. P. Barradas, M. Fonseca, J. Cruz, A. Jesus, *Eur. Phys. J. Plus* **2021**, *136*, 684.
[109] M. Razeghi, A. Rogalski, *J. Appl. Phys.* **1996**, *79*, 7433.
[110] J. Zaumseil, H. Sirringhaus, *Chem. Rev.* **2007**, *107*, 1296.

# Supporting Information

1. Contact optimization

Figure S1 shows the fits to the *I-V* curves obtained for an MSM structure device as deposited and after annealing at temperatures from 200 to 800 °C. The experimental data is also shown in Figure 2a, and the effective Schottky barriers are plotted in Figure 2b.

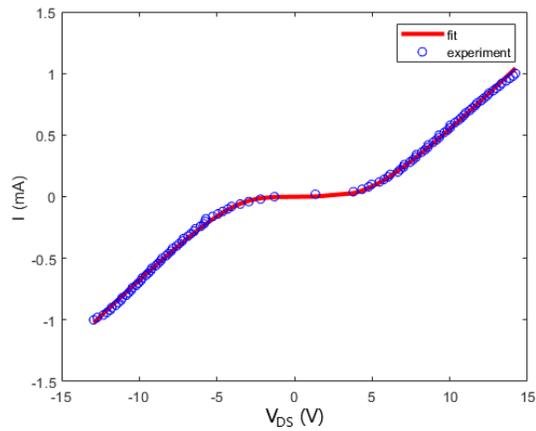

As dep.; $q\varphi_1$=0.69 eV, $q\varphi_2$=0.65 eV, $R_{membrane}$=7.59 kΩ

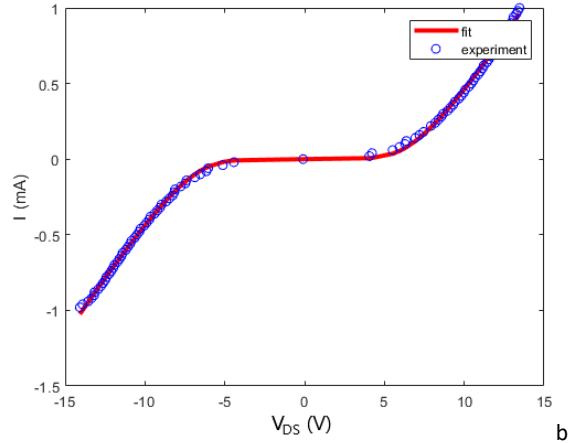

200 °C; $q\varphi_1$=0.74 eV, $q\varphi_2$=0.75 eV, $R_{membrane}$=5.76 kΩ

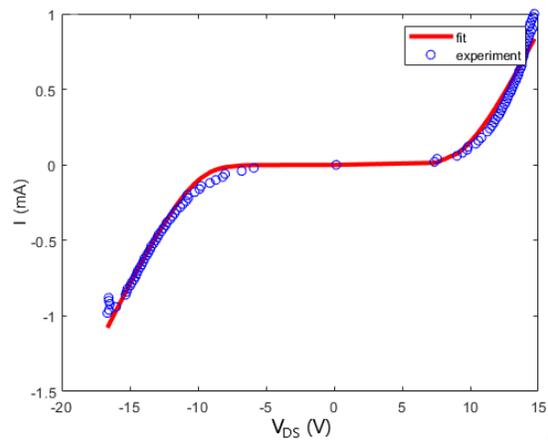

300 °C; $q\varphi_1$=0.84 eV, $q\varphi_2$=0.85 eV, $R_{membrane}$=4.79 kΩ

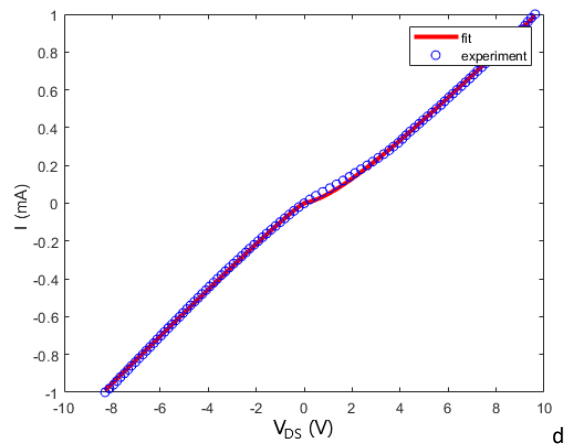

400 °C; $q\varphi_1$=0.54 eV, $q\varphi_2$=0.46 eV, $R_{membrane}$=7.58 kΩ

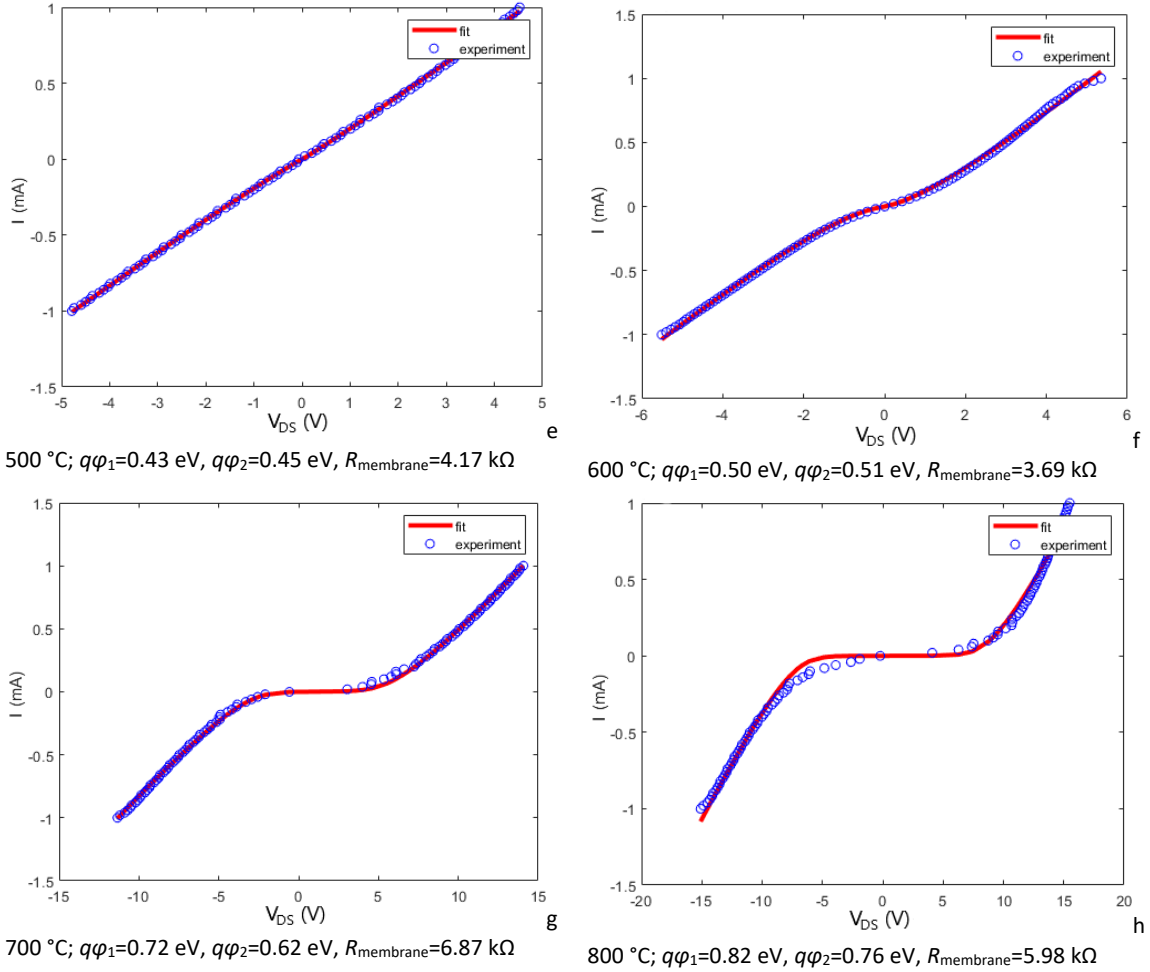

500 °C; $q\varphi_1$=0.43 eV, $q\varphi_2$=0.45 eV, $R_{membrane}$=4.17 kΩ

600 °C; $q\varphi_1$=0.50 eV, $q\varphi_2$=0.51 eV, $R_{membrane}$=3.69 kΩ

700 °C; $q\varphi_1$=0.72 eV, $q\varphi_2$=0.62 eV, $R_{membrane}$=6.87 kΩ

800 °C; $q\varphi_1$=0.82 eV, $q\varphi_2$=0.76 eV, $R_{membrane}$=5.98 kΩ

*Figure S1 – Graphic representation of the PKUMSM fits done to the experimental I-V points for the device as deposited and after each annealing step.*

2. High detectivity MSM photodetector

While in the main text, device PD-A is explored, in this supporting information, we provide a quick discussion of an MSM device with a higher detectivity, device PD-B. In Figure S2, *I-V* curves for both these devices are shown in the dark and under 245 nm wavelength illumination. Under illumination, the current is higher for device PD-A, which leads to its higher *R* of 2.6x10$^4$ A/W, compared to the 3.3x10$^3$ A/W of device PD-B at the same –40 V. However, device PD-B has a much lower dark current, particularly for negative bias. *D\** is particularly sensitive to $I_{dark}$, so the detectivity of device PD-B is much higher. Figure S3 compares the two devices in terms of *R*, *PDCR*, and *D\**, for values of bias between -40 and 40 V. Once again, an enormous difference is found in for negative bias, with device PD-B showing a much higher *PDCR* of 2.6x10$^7$ and *D\** = 2.6x10$^{16}$ Jones at *V* = −40 V, compared to the *PDCR* = 2.9x10$^2$ and *D\** = 2.4x10$^{14}$ Jones of device PD-A.

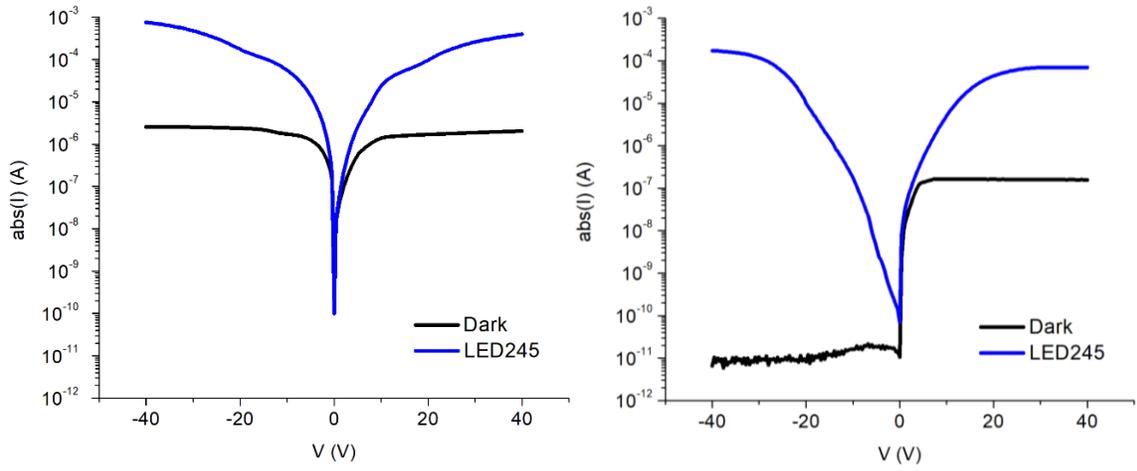

*Figure S2 – I-V curves obtained in logarithmic scale in the dark and under illumination, for **a** – device PD-A; **b** – device PD-B.*

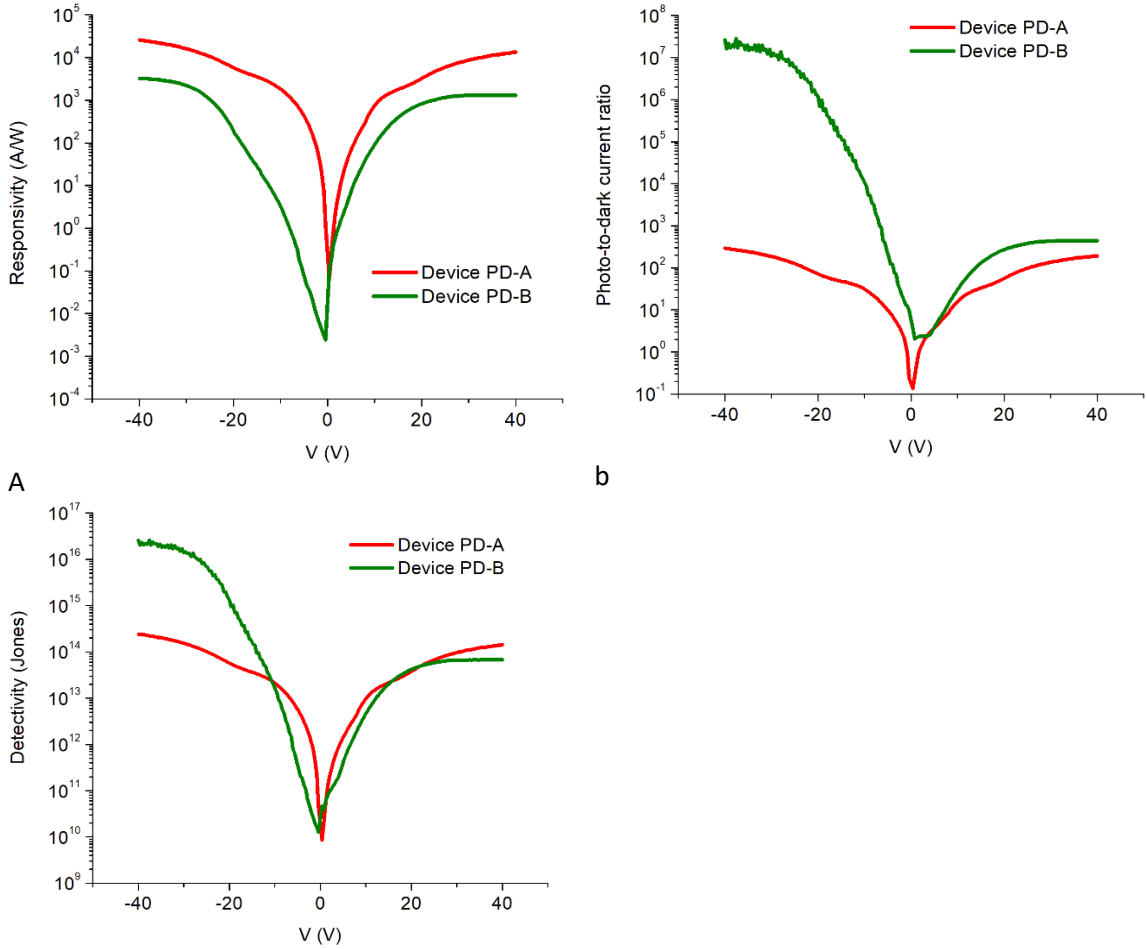

*Figure S3 – Plots, as a function of the bias applied between the contacts of devices PD-A and PD-B, of the **a** – responsivity; **b** – photo-to-dark current ratio; **c** – detectivity.*